\documentstyle[12pt,aps,epsfig]{revtex}
\tightenlines
\begin{document}
\def\ii{\'{\i}}
\def\bi{\bigskip}   
\def\be{\begin{equation}}
\def\en{\end{equation}}
\def\bq{\begin{eqnarray}}
\def\eq{\end{eqnarray}}
\def\noi{\noindent}
\def\bc{\begin{center}}
\def\ec{\end{center}}
\def\beit{\begin{itemize}}
\def\eit{\end{itemize}}
\def\ct{\centerline}
\def\tr{{\rm tr}}
\title{Unstable particles, gauge invariance and the $\Delta^{++}$
resonance parameters}  
\author{ Gabriel L\'opez Castro$^1$ and Alejandro Mariano$^2$ \\
{\small\it $^1$ Depto. de F\'\i sica, Cinvestav del IPN Apdo. Postal
14-740, 07000 M\'exico DF, M\'exico \\
$^2$ Depto de F\'\i sica, Fac. Cs. Exactas, Universidad Nal. de la Plata,
1900  La Plata, Argentina}}
\maketitle

\begin{abstract}
The elastic and radiative $\pi^+p$ scattering are studied in the framework
of an effective Lagrangian model for the $\Delta^{++}$ resonance and its
interactions. The finite width effects of this spin-3/2 resonance are
introduced in the scattering amplitudes through a complex mass scheme to
respect electromagnetic gauge invariance. The resonant pole
($\Delta^{++}$) and
background contributions ($\rho^0,\ \sigma,\ \Delta$ and neutron states)
are separated according to the principles of the analytic S-matrix theory.
The mass and width parameters of the $\Delta^{++}$ obtained from a fit to
experimental data on the total cross section are in agreement with the
results of a model-independent analysis based on the analytic S-matrix
approach. The magnetic dipole moment
determined from the radiative $\pi^+p$ scattering is
$\mu_{\Delta^{++}}=(6.14\pm 0.51)$ nuclear magnetons.
 \end{abstract}
PACS: 13.75 Gx, 14.20.Gk, 13.40.Em

\maketitle
\section{Introduction}
\label{sec:intro}

  Elementary particles with spin larger than 1 have not been discovered
yet. However, composite higher spin particles have been observed in
nature as bound states of quarks.$^1$ On the other hand, the
formulation of a fully consistent quantum field theory for these particles
is far from being complete. Thus, the description of the dynamics of such
hadronic particles is usually done in terms of an effective Lagrangian
model. Such relativistic models of classical fields are built using as a
guide the relevant symmetries underlying the dynamics of the specific
higher spin particles. Their use for phenomenological purposes remains 
consistent as long as we restrict to a tree-level description of the
amplitudes for physical processes. 

Here, we consider the case where spin 3/2 particles are unstable. To be
more specific, we are
interested in the case of the $\Delta(1232)$ baryon resonance and in the
way we introduce its finite width
effects in their associated physical observables without destroying the
symmetries of the effective Lagrangian (particularly, the electromagnetic
gauge invariance and the invariance under contact transformations).
  Our goal is to provide a framework where the intrinsic properties of
this particle, such as the  mass, width, and magnetic dipole moment
can be determined from experimental data in a consistent and well defined 
way. By this we mean that those properties share good physical
requirements such as model independence (whenever it becomes possible),
unitarity, independence upon {\it ad hoc} form factors, and
invariance under the relevant symmetries of the interactions.


The $\Delta^{++}$ resonance has spin J=3/2 and isospin I=3/2. In some of
the most popular Lagrangian formulations, its dynamics can be described in
terms of a  Rarita-Schwinger field $\psi^{\mu}(x)$. The dynamics of its
interactions with pions, nucleons and the electromagnetic field is
governed by an {\it effective} non-renormalizable Lagrangian.$^2$
Most of the problems related to the quantization of the free and the 
interacting theory of spin 3/2 particles$^3$  (see the
invited talk of Prof.
Sudarshan at this meeting)  are absent when we use the Feynman rules to
compute 
amplitudes only at the tree-level, as it will be the case in the present
work.  For the purposes of this work, such an effective Lagrangian must
be able to describe the production and decay of the $\Delta^{++}$
resonance in the elastic and radiative $\pi^+p$ scattering. As in any
effective theory of  the strong and electromagnetic interactions,
the physical (S-matrix) amplitudes derived from our Lagrangian must
be invariant under strong isospin and electromagnetic gauge
transformations. Furthermore, this model must be invariant also under  
the so-called {\it contact} transformations.$^4$ The contact
transformations are necessary to eliminate the unphysical 
components from the {\it on-shell} 
spin 3/2 fields. This does not prevent that the propagation of a virtual
spin-3/2  resonance carries spin-1/2 components that contribute to the
physical amplitude.$^5$ An important ingredient of
our model is to use a recipe$^2$
to incorporate the finite width effects of the $\Delta^{++}$ resonance
into the amplitude without spoiling the invariance under the above
symmetries that is respected in the case of stable spin 3/2 particles.

   A few more remarks are in order. The determination of the mass, width,
and magnetic dipole moment (MDM) of the $\Delta^{++}$ resonance have been
considered by many authors in the past (see the Particle Listings in 
Ref.~ [1]). Concerning the determination of
the mass and width parameters, the definitions used by authors falls into
two categories: the {\it conventional} and the {\it pole} parameters.$^6$ 
In the conventional approach, these resonance
parameters are determined
by applying the method of partial-wave analyses that use generalized
Breit-Wigner formulae to fit the experimental data. This definition of
mass and width have a strong model dependence as far as
each group has its own prescription for the treatment
of analiticity, the choice of background and the particular
parametrization of the Breit-Wigner formula.$^6$ Furthermore,
an unambiguous
separation of resonance and background contributions in this case is hard
to accomplish. In contradistinction, in the pole of the S-matrix approach
the pole position is a physical property (a process- and model-independent
property) of the S-matrix amplitude. Thus, the mass and width parameters
of a resonance can be defined from this pole position in a more 
satisfactory way. It is worth to mention that the numerical values of the
mass and width of the $\Delta^{++}$ defined from the pole position are
significantly smaller$^7$ (1.7\% and 15\%, respectively for
the mass and width) than their counterparts in the conventional approach. 

  On another hand, the determination of the magnetic dipole moment of a
resonance is necessarily model-dependent since one is forced to specify
the photon couplings to other particles. 
Different prescriptions to enforce gauge invariance, to
incorporate the resonance character of the $\Delta^{++}$, and to introduce 
 other structure-dependent effects (for example, some ad-hoc form
factors) usually lead to different results for the MDM even using the
same
experimental data.$^1$ Given this diversity of theoretical methods
and approximations, the PDG$^1$ prefers to quote an estimate for
the magnetic
dipole moment in the (rather wide) range $3.5 \leq \mu_{\Delta^{++}} \leq
7.5$ (in units of nuclear magnetons). 

   The issue of gauge invariance for processes involving unstable
particles has received great attention in the last years and deserves a
separate comment. This was
motivated by the necessity to have a consistent definition of mass
and width for the $Z^0$ (and $W$) gauge boson in view of the very precise
measurements carried out at LEP. More precisely, R. Stuart has pointed
out in the early nineties that the definition of the mass of the $Z^0$
gauge boson in the on-shell scheme was not gauge-invariant.$^8$
He has
proposed to carry out a Laurent expansion of the full
(calculated perturbatively) amplitude of 
$e^+e^- \rightarrow f\bar{f}$ around the pole position and separate the
amplitude into resonant and background term.$^8$ The pole and
background terms in the amplitude resulting from this expansion are
separately gauge-invariant, and can provide a gauge-invariant definition 
for the mass and width of the $Z^0$ boson. This pole + background
structure of  the amplitude is the same as the one imposed by general
arguments of 
the analiticity of the S-matrix that involves the production and decay of
resonances.$^9$ Later, using the Pinch Technique, Pilaftsis and
Papavassiliou$^{10}$ were able to obtain an unstable propagator
which provided 
a definition of mass and width parameters of a gauge-boson resonance
satisfying good physical properties and, in particular, gauge invariance.

In two recent papers$^{11}$ we have extended these ideas to the
sector of
the baryonic $\Delta^{++}$ resonance. Using an effective Lagrangian  
model to describe the $\Delta^{++}$ and its interactions with the $\pi^+,\
p$ and the photon fields, we have been able to incorporate the finite
width effects of this resonance without spoiling the
symmetries of the model that are satisfied in the case of
the zeroth-width approximation. In addition,
the background contributions that originate from the exchange of 
intermediate states other than the $\Delta^{++}$ (namely, the $\rho^0,\
\sigma,\ \Delta^0$ and the neutron states) are also incorporated in our
effective Lagrangian model. The full amplitude obtained in our approach
has the pole plus background structure dictated by the analytic S-matrix 
theory.$^9$ Each one of these terms in the amplitudes are
separately gauge-invariant and the insertion of ad hoc form factors to
restore gauge invariance is not necessary in our case.  

In this talk we summarize the main aspects of our analysis.  We emphasize
from our results the model-independent aspects of the mass, width and
magnetic dipole moment parameters that follows from our separation of pole
and background contributions. In a first step, we fix the mass, the width
and the strong coupling of the $\Delta^{++}$ from the elastic $\pi^+ p$
scattering. Then we obtain the $\Delta^{++}$ MDM from the radiative
$\pi^+p$ scattering observables. It is interesting to note that the 
elastic scattering requires the contribution of the scalar $\sigma$
meson in the $t$-channel to get an improved fit of the data. The details
of the different calculations and input data can be found in 
Ref.~[11].

 \section{The effective Lagrangian}
\label{sec:method}
  In this section we provide the pieces of the Lagrangian for the 
Rarita-Schwinger field $\psi^{\mu}(x)$ that are relevant to describe the 
$\Delta^{++}$ contributions to the elastic ($\pi^+ p \rightarrow \pi^+ p$)
and radiative ($\pi^+ p \rightarrow \pi^+ p \gamma$) processes of our
interest. The interaction Lagrangians for $\rho^0,\ \sigma$ mesons and the
neutron intermediate states that contribute to the background amplitude 
are well known and can be found for example in Ref.~ [11].

  The Lagrangian that describes the $\Delta^{++}$ and its interactions
with the pion ($\phi$), proton ($\psi$) and photon ($A_{\mu}$) fields is
given by:
\be
{\cal L}_{\Delta}={\cal L}_0+{\cal L}_{\Delta\pi p}+{\cal L}_{\Delta   
\Delta \gamma} + {\cal L}_{\Delta\pi p \gamma}.
\en  
The different pieces in this Lagrangian have explicitly forms:
\bq
{\cal L}_0&=&\overline{\psi}^{\mu} \Lambda_{\mu \alpha}(A)
G^{\alpha \beta} \Lambda_{\beta \nu}(A) \psi^{\nu}\ , \\
{\cal L}_{\Delta\pi p}&=& \left( \frac{f_{\Delta N \pi}}{m_{\pi}} \right)
\overline{\psi}^{\mu} \Lambda_{\mu \nu}(A) \psi \partial^{\nu} \pi +
h.c. \\
{\cal L}_{\Delta
\Delta \gamma} &=&-2e \overline{\psi}_{\nu} \Lambda^{\nu \nu'}(A)
\Gamma_{\nu' \mu' \alpha} \Lambda^{\mu' \mu}(A) \psi_{\nu}
A^{\alpha}\ , \\
{\cal L}_{\Delta \pi p \gamma}&=& \left( \frac{ef_{\Delta N \pi}}{m_{\pi}}
\right) \overline{\psi}^{\mu} \Lambda_{\mu \nu}(A) \psi \pi
A^{\nu} + h.c.
\eq
The rank two tensors introduced in the above Lagrangians are defined as
follows:
\bq
G^{\alpha \beta} &\equiv & g^{\alpha \beta}(i\! \! \not
\partial-M)+\frac{i}{3}
(\gamma^{\alpha}\! \! \not \partial
\gamma^{\beta}-\gamma^{\alpha}\partial^{\beta}-\partial^{\alpha}
\gamma^{\beta}) + \frac{1}{3}M\gamma^{\alpha}\gamma^{\beta}, \\
\Lambda_{\mu \nu}(A)\!\! &\equiv & g_{\mu \nu}
+\frac{1}{2}(1+3A)\gamma_{\mu}\gamma_{\nu}\ .
\eq
We have defined the electromagnetic vertex of the $\Delta^{++}$ following 
Ref.~[2]:
\be
\Gamma_{\alpha \beta \rho}\!\! =\! \!\left(
\gamma_{\rho}-\frac{i\kappa_{\Delta}}{2M}\sigma_{\rho
\sigma}k^{\sigma} \right)g_{\alpha \beta} -\frac{1}{3} \left(
\gamma_{\rho}\gamma_{\alpha}\gamma_{\beta}+   
\gamma_{\alpha}g_{\beta \rho}-\gamma_{\beta}g_{\alpha \rho} \right)\ ,
\en
where the $\Delta^{++}$ MDM is given by:
\be
\mu_{\Delta^{++}}= 2(1+\kappa_{\Delta}) \frac{e}{2M}\ 
\en
and  $\kappa_{\Delta}$ is the {\it anomalous} part of the magnetic dipole 
moment.  In the above equations, $m_{\pi}$ and $M$ denote the pion and
$\Delta^{++}$ masses while $f_{\Delta N\pi}$ is the (strong) coupling
constant of the $\Delta N \pi$ vertex. The isospin-invariant version of
the
above Lagrangian can be found in Ref.~[11]. The Feynman rules
associated to these Lagrangian can be found in Ref.~[2].

  The arbitrary parameters $A$ that appears in the tensor $\Lambda_{\mu
\nu}(A)$ is associated to the contact transformations acting upon the
Rarita-Schwinger field ($a\not = -1/4$):
\be
\psi^{\mu} \rightarrow \psi^{\mu}+a\gamma^{\mu} \gamma_{\alpha}
\psi^{\alpha}, \ \ A\rightarrow A'=\frac{A-2a}{1+4a}\ . 
\en
One can easily prove that the above (free and interacting) Lagrangian
remains invariant under
these contact transformations. As is was proven explicitly for the
case of elastic
and radiative $\pi^+p$ scattering,$^2$ the S-matrix  amplitudes
for these
processes are {\it independent} of the arbitrary parameter $A$ as it
should be. Furthermore, the following Ward identity between the
propagator $P^{\mu \nu}(p)$ (see Ref.~[2]) and the
electromagnetic vertex  of the $\Delta^{++}$
(see eq. (8)) \be
P^{\mu \alpha}(P') \Gamma_{\alpha \beta \rho} k^{\rho} P^{\beta \nu}(P) =
P^{\mu \nu}(P)- P^{\mu \nu}(P')\ ,
 \en
assures that the S-matrix amplitude of the radiative $\pi^+p$ process is
also gauge-invariant. 

   In summary, the model for the $\Delta^{++}$ and its interactions with
other particles described in this section give rise to S-matrix amplitudes
which are gauge-invariant and satisfy invariance under contact
transformations. This conclusion holds as far as we consider the $\Delta$
as an stable particle. Introducing the decay width naively in the
denominator of the propagator destroys gauge invariance. In the
following section we discuss a mechanism to introduce consistently the
finite width effects.

 \section{Recipe for unstable particles}
\label{sec:2point}
  Consider a radiative processes that is dominated by the production of a 
resonance in the  $s$-channel. Using a propagator of an unstable as it is
obtained from Dyson summation of bubble graphs and the on-shell
renormalization scheme leads to an  amplitude that is not invariant under
electromagnetic gauge transformations. The radiative amplitude can be
rendered gauge invariant if we replace $m_0^2 \rightarrow m^2-im\Gamma$ in
all the Feynman rules of the model, where $m_0$ is the
bare mass of the particle and $m\ (\Gamma)$ is its physical mass (width).
This {\it complex mass} recipe was proposed as a solution to 
recover electromagnetic gauge invariance of the amplitude of resonant 
processes in Ref.~[12].

   To illustrate the origin of this recipe, let us consider a resonant
scalar particle as a simple example. As is well known, the
self-interactions of this particle during his propagation transforms its 
bare propagator
\[
D_0(q^2)=\frac{i}{q^2-m_0^2} \]
into the renormalized propagator
\be
 D(q^2)=\frac{i}{q^2-m^2-iZ\mbox{\rm Im}\Pi(q^2)} \ ,
\en 
if we use  the renormalization conditions $m_0^2 = m^2-\mbox{\rm
Re}\Pi(m^2),\
Z^{-1}=1-\mbox{\rm Re}\Pi'(m^2)$. In this definition, $m$ becomes the
renormalized mass. The unitarity condition of the S-matrix
amplitude allows to identify $Z\mbox{\rm
Im}\Pi(q^2)=- \sqrt{q^2}\Gamma(q^2)$, where $\Gamma(q^2)$ is the decay
width
of the scalar particle with (virtual) mass $q^2$. 

   Let us consider now this renormalized propagator in a physical
process. One of the simplest radiative process is the scattering reaction 
$\pi^+(p) \eta(q) \rightarrow \pi^+(p')\eta(q') \gamma(k,\epsilon)$ which
we assume to be 
dominated by the production of the charged scalar resonance $a_0^+$ in the
$s$-channel (letters within parenthesis denote the four-momenta and
$\epsilon$ the photon polarization vector). There are three resonant
contributions corresponding to the photon emitted from the external pion
lines and from the internal $a_0^+$ propagator line. The transition
amplitude is
given by: \be
{\cal M}= eg^2 \left\{- \frac{p.\epsilon}{p.k} D(Q') +
\frac{p'.\epsilon}{p'.k} D(Q) -iD(Q)D(Q')(Q+Q').\epsilon \right\}\ .
\en
We have introduced the variables $Q=p+q,\ Q'=p'+q'$ ($Q=Q'+k$) which
denote the four-momenta of the intermediate $a^+_0$ resonance, and
$D(Q_i)$ denote its resonant propagator as given in eq. (12). The factor
$g$ denotes the coupling constant for the $a_0 \eta \pi$ vertex.

  We can easily check that the above amplitude is not gauge-invariant,
namely that ${\cal M}$ does not vanish when $\epsilon$ is replaced by
$k$.
Gauge
invariance is not satisfied due to the presence of the
(energy-dependent) imaginary part of the propagator. Gauge
invariance can be restored in different forms, introducing in this way an
ambiguity in the amplitude. One can for instance include form factors
in the strong vertices or additional contributions to the amplitude in an 
{\it ad hoc} way. A second possibility is to include the one loop
corrections  to the electromagnetic vertex of the $a^+_0$ meson in order
to satisfy a Ward identity at the one-loop level. A third option consists
in using a complex mass scheme as proposed in Ref.~[12]. 

   We consider here the complex mass scheme  since it provides the
simplest solution. For the illustrative example under consideration, let
us
consider only the absorptive corrections to the propagator. If we assume a
renormalized mass for the $a_0$ from the beginning, we can write the 
absorptive part of the self-energy correction as follows:
\be
- \mbox{\rm Im}\Pi(s) =\sqrt{s}\Gamma_{a_0}(s)=\frac{g^2}{16\pi s} \left\{
(s-(\mu+\mu')^2)(s-(\mu-\mu')^2 \right\}^{1/2}\ ,
 \en
where $\mu\ (\mu')$ denotes the mass of the $\eta\ (\pi^+)$ meson running
in the loop correction. In the limit of {\it massless} particles in the
loop
we can check that:
\be
-\mbox{\rm Im}\Pi(s) \rightarrow \frac{g^2}{16\pi} = M\Gamma\ ,
\en
where $\Gamma$ is the decay width of the $a_0^+$ meson and $M$ its mass.
Thus, the resonant propagator becomes:
\be
D_{a_0}(s)  \rightarrow \overline{D}_{a_0}(s) =\frac{i}{s-M^2+iM\Gamma}\ .
\en
This propagator can be obtained from the bare propagator if we simply
replace the bare mass by the pole position $M^2-iM\Gamma$, namely if we
adopt the complex mass scheme. Owing to the identity:
\be
\overline{D}_{a_0}(Q') \overline{D}_{a_0}(Q) =\frac{i}{(Q+Q').k} \left\{
\overline{D}_{a_0}(Q')-\overline{D}_{a_0}(Q) \right\}\ ,
\en
we can check that using the resonant propagator (16) in the limit of
massless loop corrections, the amplitude eq. (13) becomes gauge
invariant.

  This simple example illustrates that the complex mass scheme and the
absorptive one loop corrections to  the electromagnetic vertex and the
propagator (in the limit of massless particles    
running in loop corrections$^{13}$) are equivalent methods
to restore gauge invariance. 
 Although this approximation (massless particles in loop corrections) can
hardly be justified in the case of the $\Delta^{++}$ resonant propagator
(because
the proton in
the loop is not massless in the chiral limit), the complex mass scheme
provides the simplest solution to the gauge invariance problem for
resonant amplitudes and it will be adopted here for our calculations.

\section{Fitting experimental data}
\label{sec:3point}
 Just to clarify our procedure, we repeat here the main steps of our
analysis. First, we use the experimental data on the total cross section
of $\pi^+p$ elastic scattering to fix some relevant free parameters (mass,
decay width and strong coupling of the $\Delta$) of the model. Then,
we use the data on radiative $\pi^+p$ scattering to fix the MDM of the
$\Delta^{++}$ which remains as the only free parameter in this reaction.
The details of the fit procedure and further additional tests of the model  
can be found in Ref.~[11]. Here we focus on the discussion of the
relevant features of the model and the results of the fit.

\subsection{Elastic $\pi^+p$ scattering}
The model contributions to the $\pi^+p \rightarrow \pi^+p$ scattering
includes the $\Delta^{++}$ resonance ($s$-channel), the $\rho$ and
$\sigma$ mesons ($t$-channel) and the $\Delta^0$ and neutron states
(crossed-channel) contributions. There are five Feynman diagrams
corresponding to these contributions which can be found in Ref.~[11]. 
The experimental data for the total cross
section is chosen to lie in the resonance region, which corresponds to
pion kinetic energies $T_{lab}=75 \sim 300$ in the lab system.$^{14}$ 
In this kinematical region, the elastic scattering is
dominated by the production of the $\Delta^{++}$ resonance and all 
other terms can be considered as small background contributions. This is
indeed the case, as it can be checked from Figure 1.

   The parameters entering the background contributions (except the
$\Delta^0$ mass and the couplings of the scalar meson) are taken from 
other low energy processes (see Ref.~ [11]). Their precise values are
not of critical importance as
far as they contribute as a small term to the amplitude. Therefore, the
only free parameters of the model are the mass ($M_{\Delta}$), width
($\Gamma_{\Delta}$) and $\Delta N \pi$ ($f_{\Delta N \pi}$) coupling of
the $\Delta$ and the effective coupling of the scalar meson
($g_{\sigma}=g_{\sigma \pi \pi}g_{\sigma NN}$). 

   In order to assess the influence of the different background terms, we
have performed several fits to the total cross section by adding
successively the different background contributions. For the mass of the
scalar $\sigma$ meson we
have chosen $m_{\sigma}=650$ MeV. We have allowed a wide variation of
this mass, namely $\Delta m_{\sigma}=\pm 200$ MeV, and have found that it
is correlated mainly with the value of $g_{\sigma}$, while the other
parameters are not affected in an important way. The results of the fit
are shown in Table 1 and in Figure 1. \\

A few comments are in order:\\

\noindent
$(i)$ The $\chi^2$/dof drops from 121 to 10 in going from the top to the
bottom of Table 1, which indicates the necessity of including in the fit
some degrees of freedom other than the $\Delta$ resonance. The large 
contributions to the $\chi^2$/dof come from the data points for the
highest pion energies considered in the fit (see Figure 1). In Figure 1 we
can observe the best results obtained for each case indicated in Table 1.
Although the $\chi^2$/dof is not indicative of a very good fit, we can
expect that such a fit can be improved by considering effects
of rescattering and other background terms excluded from the simple pole
approximation implicit in our model.

$(ii)$ Since only the amplitude involving
the $\Delta^{++}$ resonance has an imaginary part, it can be easily
checked that the
complete amplitude does not satisfy unitarity. We can force our result to
satisfy unitarity by adding a softly energy-dependent term to the 
amplitude. The presence of additional terms in the amplitude can be  
justified on the basis that we have kept only the pole term in our
amplitude for the $\Delta^{++}$ contribution.$^{11}$ The result
obtained from the fit when we impose unitarity is shown in the fifth row
of Table 1. Namely, unitary only shifts the value of the decay width
 in the right direction to match the model-independent result$^7$
 shown in the last row of Table 1.

$(iii)$ The row denoted as `pole position' in Table 1 contains the results
of the fit obtained in a model-independent analysis of the same data for
the cross section.$^7$ The close agreement observed in Table
1 between our model-dependent results and the model-independent analysis
of Ref.~[7] indicates that our model describes very well
not only the resonant but also the background contributions in the
amplitude. 

$(iv)$ Once the relevant parameters of the model are fixed from the total
cross section, we can {\it predict} the differential cross section
$d\sigma/d\Omega$ for pions emitted at an angle $\theta$ with respect to
the incident pion beam. Our model is able to reproduce two sets of data
corresponding to kinetic energies of incident pions at $T_{lab}=263$ and
$291$ MeV.$^{11}$ The test of the model at these energies is important 
because the data used to extract the magnetic dipole moment of the 
$\Delta^{++}$ correspond to kinetic energies close to those values (see
next section).

  These important remarks indicates that our model is well suited to
describe the dynamics of the $\pi^+p$ reactions in the
$\Delta^{++}$ resonance region and provides good confidence to apply it in
the description of other reactions. In the next section, we use it to
extract the MDM of this resonance from the data on radiative $\pi^+p$
scattering.

\subsection{Radiative $\pi^+p$ scattering}
Once we have fixed the mass and width of the $\Delta$ and other relevant
couplings of the model, it remains only one parameter to describe the
radiative $\pi^+p$ scattering: the magnetic dipole moment of
$\Delta^{++}$, namely $\mu_{\Delta^{++}}$ or $\kappa_{\Delta}$. In this
section, we fit this parameter from experimental data on $\pi^+p
\rightarrow \pi^+ p \gamma$.
From the  35 Feynman diagrams (see Ref.~ [11]) that contribute to
this process in our model, seven correspond to photons emitted from
process involving the $\Delta^{++}$ intermediate states and 28 are
associated with the $\rho^0,\
\sigma$ and $\Delta^0,\ n$ intermediate states.

   The physical observable of our interest is the five-fold differential
cross section $d\sigma/d\omega_{\gamma} d\Omega_{\gamma} d\Omega_{\pi}$ of
the radiative $\pi^+p$ scattering. In this observable, $\omega_{\gamma}$
is the photon energy, $d\Omega_{\gamma}$ is the element of solid angle
where photons are emitted with respect to final pions, and $d\Omega_{\pi}$
is the solid angle for final state pions measured with respect to the
direction  of incident pions. We use the data corresponding to incident
pions of energies $T_{lab}=269$ and $298$ MeV.$^{15}$ As we
have pointed out in the previous section, our model is still good to
describe the data at those energies. The kinematical range of
photon energies is $0 \leq \omega_{\gamma} \leq 100$ MeV, where we can
expect, in
a conservative way, that the soft-photon approximation is valid and that
other structure dependent terms or higher electromagnetic multipoles
contributions are small. In addition, we consider a few set of photon
angular configurations where the differential cross section is more
sensitive to the effect of the $\Delta^{++}$ MDM (see Ref.~ [11]).

   The results of the fits for the most sensitive observables are shown
in Table 2 (the definition of the `anomalous' part $\kappa_{\Delta}$ of
the MDM was given in Eq. (9)):

Again, a few remarks are worth to be mentioned:

$(i)$ In Figure 2 we show the best fits for a few samples of the
differential cross section as a function of the photon energies for
$T_{lab}=269$ MeV. Just to show the sensitivity of the chosen
configurations (G1, G4 and G7) to the effect of the magnetic dipole
moment, in Figure 2 we compare the best fits of Table 2 (solid lines) with
the curves
corresponding to a reference value $\kappa_{\Delta}=1$ (dashed curves). 

$(ii)$ Other (less sensitive to $\kappa_{\Delta}$) measured angular
configurations were also considered in the analysis. The description of
data is very good as it can be checked in Ref.~[11].

$(iii)$ The set of values determined for $\kappa_{\Delta}$ (see Table 2)
is remarkable consistent. This allows to quote a meaningful weighted
average from the six values of $\kappa_{\Delta}$ shown in Table 2. We
obtain:
\be
\mu_{\Delta^{++}} = 2(1+\kappa_{\Delta})\frac{e}{2m_{\Delta}} = (6.14
\pm 0.51)\frac{e}{2m_p} 
\en
Note that the last numerical value is given in units of nuclear magnetons.

\section{Remarks and Conclusions}
The contributions of the $\Delta^{++}$ resonance to the elastic and
radiative $\pi^+p$ scattering is revisited in the light of a
consistent effective Lagrangian model for the spin-3/2 {\it
unstable} particle and its interactions. Our proposal respects two
very important symmetries of a theory of the spin-3/2 particles: the
invariance under {\it contact} and electromagnetic gauge transformations.
We have shown that introducing the finite width effects of the resonance
through a complex mass scheme, namely replacing $M^2 \rightarrow
M^2-iM\Gamma$ in all the Feynman rules that involve the spin-3/2 particle,
do not spoil these symmetries of
the effective theory. Such a recipe is well motivated by recent studies
concerning the search a proper (gauge-invariant) definition of the mass of
an unstable gauge-boson in the framework of perturbative gauge theories. 

  We have performed a phenomenological analysis of this effective
Lagrangian  to test its viability as an acceptable model for the low energy
$\pi^+p$ scattering processes. Our approach is closely related to the one
of
the analytic S-matrix; namely, we try to give a physical meaning to the
mass and
width of the resonance from an explicit separation of resonant and
background contributions in the S-matrix amplitude for the elastic
scattering. By introducing the complex mass scheme in the propagator of
the resonance we are able to isolate the pole contributions in a simple
and clean way. The background contributions are given in our model by
the exchange of the $\rho^0,\ \sigma,\ \Delta^0$ and neutron
intermediate states. In the case of elastic $\pi^+p$ scattering, we have
found
that the mass and width of the $\Delta^{++}$ resonance are in excellent
agreement (within the approximations inherent to our model) with the
values obtained in the framework of the model-independent analytic
S-matrix approach (see Table 1). The description of the elastic scattering
data for the total
cross section is very good in a wide region considered around the
resonance peak. The differential cross section of elastic scattering
 is predicted to be in good agreement with a set of data  for pion
kinetic energies to the right side of the resonance peak.

  We have considered also the radiative $\pi^+p$ scattering in view of
extracting a value of the $\Delta^{++}$ magnetic dipole moment from the
experimental data. Electromagnetic gauge invariance is fulfilled for
the resonance contributions to the amplitude owing to a simple Ward
identity that is satisfied between the propagators and the electromagnetic
vertex of the $\Delta^{++}$ in the complex mass scheme. Our model provides
a very simple solution to the gauge invariance problem for the resonance 
contributions to the radiative amplitude in the presence of finite width
effects. Within our approach we do not need to introduce {\it ad hoc}
form factors or additional contributions to obtain a gauge-invariant
amplitude.

   Using the most sensitive set of data for the five-fold differential
cross section of the radiative $\pi^+p$ scattering we are able to give a
good fit with only one free parameter: the $\Delta^{++}$ magnetic dipole
moment. The results for the $\Delta^{++}$ MDM are described in section
IV.2 and can be found in Table 2 and equation (18). Our determination of
the MDM are in good agreement with recent theoretical calculations that
incorporate the QCD corrections in a chiral
bag model$^{16}$ and with the predictions of a phenomenological
quark model$^{17}$ that includes the non-static effects of pion
exchange and orbital excitation. Our determination of the MDM is, however, 
a bit larger that some calculations based on the SU(6) spin-flavor
symmetry.$^{18}$

  In summary, we have shown that the data on the elastic and radiative
$\pi^+p$ scattering near the $\Delta^{++}$ resonance region can be
well described in the framework of an effective Lagrangian model for this
spin-3/2 particle. This model is free from the very well known
inconsistencies present in the quantum field theory formulations of
spin-3/2 particles as far as we use the model only at the tree-level. The
calculations of the scattering amplitudes fully exploit the
model-independent separation of the amplitude into the resonant and
background contributions advocated by the analytic S-matrix theory. This
allows us to give a physical meaning to the mass and width values
extracted for the $\Delta^{++}$ resonance and, by extension, to its the
magnetic dipole moment parameter.

\section{Acknowledgements}

I would like to thank D. V. Ahluwalia and M. Kirchbach for their kind
invitation to this meeting. The partial financial support from Conacyt is
gratefully acknowledged. 

\section{References}

\noindent
$^1$K. Hagiwara et al, {\it Review of Particle Physics}, Phys.
Rev. {\bf D66} Part I, (2002).

\noindent
$^2$M. El-Amiri, G. L\'opez Castro and J. Pestieau, Nucl.
Phys. {\bf A543}, 673 (1992).

\noindent
$^3$K. Johnson and E. C. G. Sudarshan, Ann. Phys. {\bf
13}, 126 (1961); 

\noindent
$^4$L. M. Nath, B. Etemadi and J. D. Kimel, Phys. Rev. {\bf
D3}, 2153 (1971); R. D. Peccei, Phys. Rev. {\bf 176}, 1812 (1968); R. E.
Behrends and C. Fronsdal, Phys. Rev. {\bf 106},
277 (1958); J. Ur\'\i as, Ph. D. Thesis, Universit'e catholique de
Louvain, Belgium 1976.

\noindent
$^5$M. Benmerrouche, R. M. Davidson and N. C. Mukhopadhyay,
Phys. Rev. {\bf C39}, 2339 (1989); V. Pascalutsa and R. Timmermans, Phys.
Rev. {\bf C60}, 042201 (1999).

\noindent
$^6$ An interesting discussion about these definitions can be
found in pp. 696-698 of the 2000 Edition of the Review of Particle
Physics.

\noindent
$^7$A. Bernicha, G. L\'opez Castro and J. Pestieau, Nucl.
Phys. {\bf A597}, 623 (1996).

\noindent
$^8$R. G. Stuart, Phys. Lett. {\bf B262}, 113 (1991); {\it
ibid}
{\bf 272}, 353 (1991).

\noindent
$^{9}$
R. Peierls, in E. H. Bellami, R. G. Moorhouse (Eds.), 
{\it Proc. 1954 Glasgow Conf. on Nucl. and Meson Physics\/},
( Pergamon Press, 1955) p.~296; 
R. Eden, P. Landshoff, D. Olive and J. Polkinghorne, 
``{\it The Analytic S-matrix}" (Cambridge University Press, Cambridge,
1966); 
M. L\'evy, Nuovo Cimento {\bf 13}, 115 (1958).

\noindent
$^{10}$
J. Papavassiliou and A. Pilaftsis, Phys. Rev. {\bf D53}, 2128 (1996);
Phys. Rev. Lett. {\bf 75}, 3060 (1995).

\noindent
$^{11}$G. L\'opez Castro and A. Mariano, 
Nucl. Phys. {\bf A697}, 440 (2002);
Phys. Lett. {\bf B517}, 339 (2001).

\noindent
$^{12}$G. L\'opez Castro, J. L. Lucio and J. Pestieau, 
Mod. Phys. Lett. {\bf A6}, 3679 (1991); 
Int. J. Mod. Phys. {\bf A11}, 563 (1996).

\noindent
$^{13}$M. Beuthe, R. Gonz\'alez Felipe, G. L\'opez Castro and 
J. Pestieau, 
Nucl. Phys. {\bf B498}, 55 (1998); 
G. L\'opez Castro and G. Toledo S\'anchez,
Phys. Rev. {\bf D61}, 033007 (2000).

\noindent
$^{14}$E. Pedroni et al., Nucl. Phys. {\bf A300}, 321 (1978).

\noindent
$^{15}$B. M. K. Nefkens et al., Phys. Rev. {\bf D18}, 3911 (1978).

\noindent
$^{16}$M. I. Krivoruchenko, Sov. J. Nucl. Phys. {\bf 45}, 109 (1987).

\noindent
$^{17}$J. Franklin, Phys. Rev. {\bf D66}, 033010 (2002).

\noindent
$^{18}$M. A. B. B\'eg, B. W. Lee, and A. Pais, Phys. Rev. Lett. {\bf 13}, 514
(1964); M. A. B. B\'eg and A. Pais, Phys. Rev. {\bf 137}, B1514 (1965); G.
E. Brown, M. Rho, and V. Vento, Phys. Lett. {\bf B97}, 423 (1980).

\newpage    
\begin{center}
{\bf Table Captions}
\end{center}

Table 1: \\

\noindent
Fit results to the total cross section of $\pi^+p$
elastic scattering.\\

Table 2:\\

\noindent
Anomalous $\Delta ^{++}$ magnetic dipole moment extracted from
radiative $\pi^+ p$ scattering.

\

\

\begin{center}
{\bf Figure Captions }
\end{center}

Fig. 1:\\

\noindent
Total cross section of elastic $\pi^+p$ scattering:
comparison of model and experimental data.\\

Fig. 2\\
Differential cross section of radiative $\pi^+p$ scattering  
and best fits results for $T_{lab}=269$ MeV and three angular
configurations
of photon energies.

\newpage

Table 1:\\

\begin{center}
\begin{tabular}{|c|c|c|c|c|c|}
\hline
Intermediate state \ & $f_{\Delta N \pi}^2/4\pi\ $ &
$m_{\Delta}$ (MeV) &
\ $\Gamma_{\Delta}$ (MeV) \ & $g_{\sigma}/4\pi$ & $\chi^2$/dof\\
\hline
$\Delta^{++, 0}$ & \ 0.281$\pm$0.001 \ & \ 1201.7$\pm$0.2 \ &
\ 69.8$\pm$0.2\ 
&-- & 121.1 \\
$\Delta^{++, 0}, N$ & \ 0.331$\pm$0.003 \ & \ 1208.6$\pm$0.2\ &
\ 87.5$\pm$0.3 \ & -- & 17.6\\
$\Delta^{++, 0}, N, \rho$ & \ 0.327$\pm$0.001 \ & \ 1207.4$\pm$0.2 \ &
\ 85.6$\pm$0.3 \ & -- & 15.6 \\
$\Delta^{++, 0},N,\rho, \sigma$ \ & \ 0.317$\pm$0.003 \ & \ 1211.2$\pm$0.4
\ &
88.2$\pm$0.4 & \ 1.50$\pm$0.12 & 10.5 \\
Unitarity & \ 0.317 \ & 1211.7 & 92.2 & 1.50 & 9.8\\
\hline \hline
Pole position & \ & \ 1212.2$\pm$0.3 \ & \ 97.1$\pm$0.4 \ & & \\
\hline
\end{tabular} \\ 
\end{center}

\newpage

Table 2:\\

\begin{center}
\begin{tabular}{|c c c c c |c|}
\hline
$T_{lab}$ (MeV)&Geometry &$\theta_{\gamma}$& $\phi_{\gamma}$ &
$\kappa_{\Delta}$ & $\chi^2$/dof  \\
\hline
& G7 & \ 120$^0$ \ \ & \ \ 0$^0$ \ \ & \ \ 3.27$\pm$0.76 \ \ & 1.99\\
269& G4 & \ 140$^0$ \ \ & \ \ 0$^0$ & \ \ 3.01$\pm$0.67 \ \ & 2.48\\
&G1& \  160$^0$ \ \ & \ \ 0$^0$ \ \ & \ \ 2.74$\pm$0.87\ \  & 1.73\\
\hline
&G7& \ 120$^0$ \ \ & \ \ 0$^0$ \ \ & \ \ 3.10$\pm$0.87\ \  & 2.68\\
298 & G4 & \ 140$^0$\ \  & \ \ 0$^0$ \ \ & \ \ 2.90$\pm$0.75 \ \ & 4.75\\
&G1 &\ 160$^0$ \ \ & \ \ 0$^0$\ \  &\ \  2.61$\pm$1.00 \ \ & 1.47\\
\hline
\end{tabular} \\ 

\end{center}

\newpage

Fig. 1:\\

\begin{center}
\begin{figure}[htb]
\centerline{
\epsfig{figure=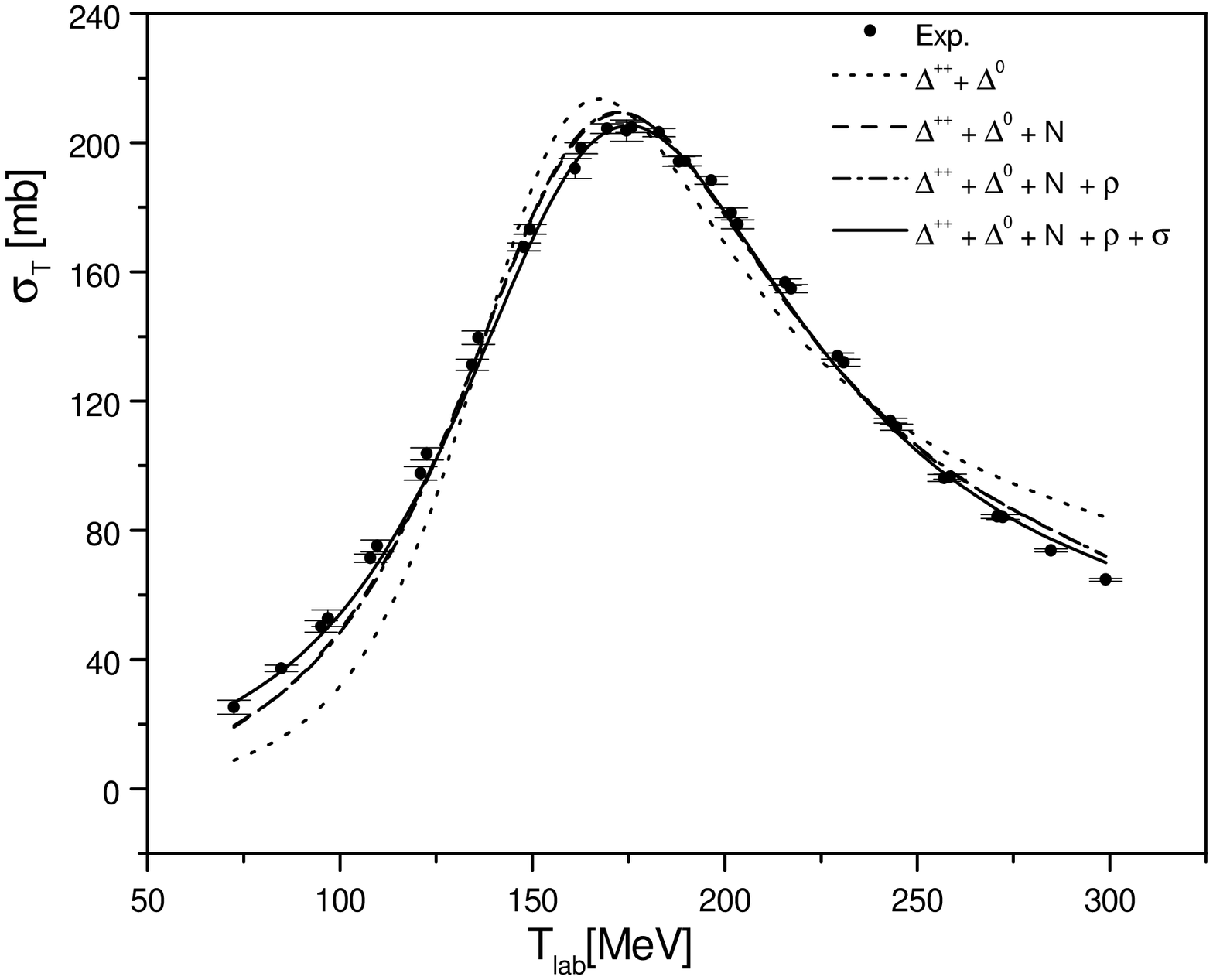,width=12.5cm}}
\vspace{-3.2cm}
\end{figure}
\end{center}

\newpage
Fig. 2:\\

\begin{figure}[htb]
\centerline{
\epsfig{figure=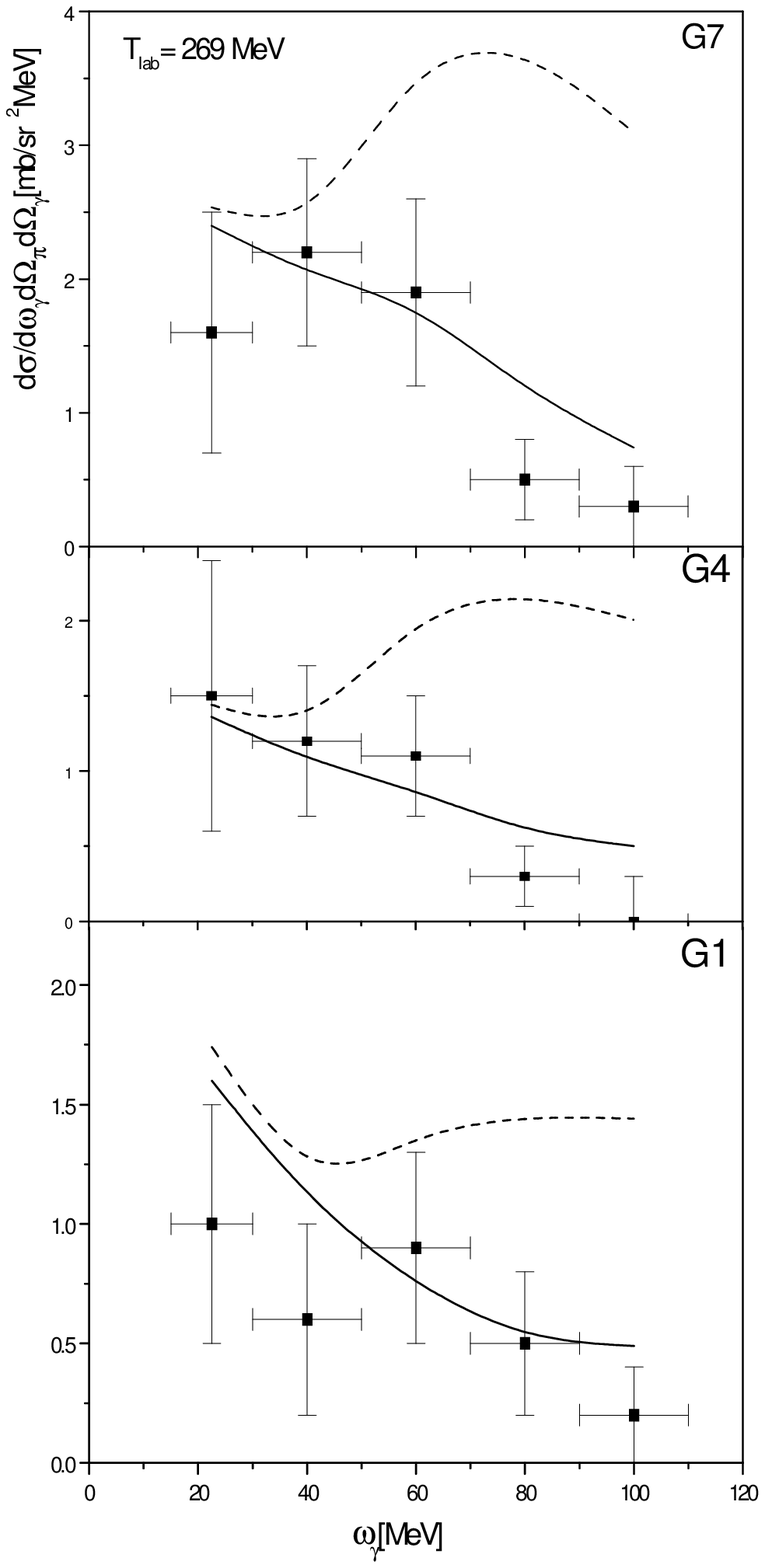,width=13.5cm}}
\vspace{-3.0cm}
\end{figure}

\end{document}